\documentclass[aps,pra,preprint]{revtex4}
\begin{document}
\baselineskip=24pt
\title{Local density approximation for exchange in excited-state density
functional theory}
\author{Manoj K. Harbola and Prasanjit Samal}
\affiliation{Department of Physics, Indian Institute of Technology, 
 Kanpur 208016,India}

\begin{abstract}

Local density approximation for the exchange energy is made for
treatment of excited-states in density-functional theory.  It is
shown that taking care of the state-dependence of the LDA exchange
energy functional leads to accurate excitation energies.

\end{abstract}

\maketitle
\newpage
Following the success of ground-state density functional theory (DFT) 
\cite{gross1,parr1}, attempts have been made to develop a similar theory for 
the excited-states. The two directions that these investigations have taken are 
based on (i) time-independent theory \cite{gorling,levy1} that is similar to the 
self-consistent Kohn-Sham formalism for the ground-state, and (ii) time-dependent
theory (TDDFT)\cite{casida,gross2} that makes use of the fact that 
frequency-dependent polarizability of a system is singular at the excitation 
frequency.  The former approach is similar to the $\Delta$-SCF method of obtaining 
transition energies; Within the ground-state DFT, a similar method was proposed by 
Ziegler et al. \cite{ziegler}  and von-Barth \cite{vonB} to calculate the energies 
of the lowest-energy multiplets.

Based on the work of Levy and Nagy, we can obtain the energy $E$ of an 
excited-state from its density $\rho({\bf r})$ from the functional
\begin{equation}
E[\rho] = \int d{\bf r}v_{ext}({\bf r})\rho({\bf r})+F[\rho,\rho_{0}]
\label{1}
\end{equation}
where $v_{ext}({\bf r})$ is the external potential and $F[\rho,\rho_{0}]$ is
a functional of $\rho$, the excited state density of a system with the 
ground-state density $\rho_{0}$. In general, however, the ground-state density
$\rho_{0}$ can be represented by the external potential itself.  Assuming
non-interacting v-representability of the excited-state density, the density
can be obtained by solving the excited-states Kohn-Sham equation
(equations are written in atomic units)
\begin{equation}
\left[-\frac{1}{2}\nabla^{2}+v_{ext}({\bf r}) + \int\frac{\rho({\bf r'})}
{|{\bf r}- {\bf r'}|}d{\bf r'}+v_{xc}({\bf r})\right]\phi_{i}({\bf r})=
\epsilon_{i}\phi_{i}
({\bf r})
\label{2}
\end{equation}
as
\begin{equation}
\rho({\bf r}) = \sum_{i}n_{i}|\phi_{i}({\bf r})|^{2},
\label{3}
\end{equation}
where $n_{i}$ is the number of electrons in orbital $\phi_{i}$.
Here $v_{xc}({\bf r})$ is the exchange-correlation potential for the 
excited-state under consideration, and is derived as the functional
derivative of the excited-state exchange-correlation energy functional.
The latter is obtained from the functional $F[\rho,\rho_{0}]$ by 
subtracting from it the non-interacting kinetic energy and the Coulomb
energy.  For details we refer the reader to the literature \cite{levy1}. 
Like the ground-state theory, in excited-state formalism also the 
exchange-correlation energy functional is not known and has to be
approximated.  However, unlike the ground-state theory, the functional is 
not universal in that it depends on the system (through $\rho_{0}$) and is
also state-dependent. Nonetheless, calculations of excitation energies
have been done \cite{gorling,harb1} employing the regular local-density 
approximation (LDA) for the ground-state. These give reasonably accurate 
excitation energies for low lying excited-states but underestimate them when 
higher excitations are considered. The purpose of this paper is to lay the 
foundations for getting the state-dependent exchange energy functional within
the local density approximation and to show that excitation energies
obtained with this functional improve over those calculated by employing
the ground-state functional for both the ground and the excited states.

We start by commenting upon why using the same (ground-state LDA functional) 
results in an underestimate of the excitation energies.  As the 
electrons are excited in a system, the overlap between the orbitals 
decreases resulting in less of exchange effect compared to the 
ground-state.  This is because now the electrons of the same spin are 
relatively less likely to come close. However, when we employ the 
ground-state functional to excited states also, this effect is ignored 
and consequently within the local approximation we make one more 
approximation.  The latter gives larger magnitude of the exchange energy 
than what the correct local approximation for the excited-states should 
give, and this results in smaller excitation energies. To elaborate on this,
let us take the example of a homogeneous electron gas. If it is in its 
ground-state, the electrons occupy wave-vectors in the k-space from $k=0$ to
$k_{F}=(3\pi^{2}\frac{N}{V})^{\frac{1}{3}}$, where $N$ is the number
of electrons distributed uniformly over a volume $V$. On the other hand, in
an excited-state of this system the electrons occupy the k-space from the 
wave-vector $k_{1}$ to $k_{2}$  so that 
\begin{equation}
k_{2}^{3}-k_{1}^{3} = 3\pi^{2}\frac{N}{V}
\label{4}
\end{equation}
The exchange energy in the latter case (the expression for the exchange 
energy is given below) is smaller in magnitude than the ground-state 
exchange energy. However, if we approximate it by the expression for the 
ground-state, its magnitude is overestimated, leading to an excited state 
energy more negative than its correct value.

As is clear from the discussion above, the correct local-density 
approximation for the inhomogeneous electron gas in an excited-state 
must be made by considering the electrons to be distributed over 
regions of k-space different from those for the ground-state. In the simplest 
case we can take the region to be a spherical shell of inner radius $k_{1}$ 
and outer radius $k_{2}$ with the two radii connected by the the relationship 
given in Eq. \ref{4} above.  Such a shell would represent an excited-state 
where the lowest energy orbitals are vacant. The expression for the 
exchange energy density for this distribution is easily derived and is given as
\begin{equation}
\epsilon_{x}=\frac{E_{X}}{V}=
-\frac{1}{8\pi^{3}}\left(2(k_{2}^{3}-k_{1}^{3})(k_{2}-k_{1})
+(k_{2}^{2}-k_{1}^{2})^{2}ln\left(\frac{k_{2}+k_{1}}{k_{2}-k_{1}}\right)
\right)
\label{5}
\end{equation}
Now for a given inhomogeneous electron gas of excited-state density 
$\rho({\bf r})$, the LDA is made by assigning two ${\bf r}$-dependent wavevectors 
$k_{1}$ and $k_{2}$ related through Eq. \ref{4} above and calculating the
exchange energy per unit volume at that point from Eq. \ref{5}. For $k_{1}=0$,
the expression above gives the ground-state LDA energy functional 
\begin{equation}
E_{x}[\rho] = -\frac{3}{4}\left(\frac{3}{\pi}\right)^{\frac{1}{3}}
\int\rho^{\frac{4}{3}}({\bf r})d{\bf r}
\label{nn1}
\end{equation}

Although the focus above has been on the exchange energy, dramatic 
effects of occupying the same region of k-space for both the ground- and
the excited-states are seen when we compare the exact non-interacting
kinetic energy and its local-density counterpart - the Thomas-Fermi
kinetic energy \cite{gross1,parr1} - for a set of orbitals occupied in the 
ground- and an excited-state configuration. For a given set of occupied 
orbitals $\{\phi_{i}\}$ with occupation numbers $\{n_{i}\}$, the former is given as
\begin{equation}
T_{s}=\sum_{i}n_{i}<\phi_{i}|-\frac{1}{2}\nabla^{2}|\phi_{i}>
\label{nn2}
\end{equation}
whereas the Thomas-Fermi kinetic energy is 
\begin{equation}
T_{s}^{TF}[\rho]=\frac{3}{10}\left(3\pi^{2}\right)^{\frac{2}{3}}
\int\rho^{\frac{5}{3}}({\bf r})d{\bf r},
\label{nn3}
\end{equation}
where $\rho({\bf r})$ is given by Eq. \ref{3}.  As an example, consider the 1s, 
2s and 3p orbitals for the $Be^{2+}$ ion occupied in different 
configurations. In the ground-state configuration ($1s^{2}$), the exact 
kinetic energy is $13.2943$ a.u. whereas the Thomas-Fermi functional gives
it to be $12.0360$ a.u. - an error of $9.5\%$.  On the other hand, if
we take the virtual orbitals $2s$ and $3p$ to be occupied with one 
electron each, the kinetic energy comes out to be $1.2381$ a.u.
whereas the Thomas-Fermi functional now gives the kinetic energy to
be $0.3090$ a.u. - an error of about $75\%$! (This is for the 2s and 3p
orbitals taken from the ground-state calculation; if we perform a
self-consistent LSD calculation with these orbitals occupied, the
answers are $2.5481$ a.u. and $0.6163$ a.u., respectively.  The
error again is about $75\%$). The error for the excited-state becomes
much larger because in calculating the Thomas-Fermi kinetic energy for
the excited-state as we are still occupying the k-space from $k_{1}=0$ to 
$k_{2}=(3\pi^{2}\rho)^{1/3}$. Better estimates of kinetic-energy via
the Thomas-Fermi approach would be obtained if we instead consider the 
electrons to be occupying a shell of inner radius $k_{1}$ and outer radius 
$k_{2}$.  In the latter case the Thomas-Fermi kinetic energy density $\tau$ 
is given as 
\begin{equation}
\tau = \frac{k_{2}^{5}-k_{1}^{5}}{10\pi^{2}}
\label{n1}
\end{equation}
or its spin-polarized version \cite{gross1,parr1}. For $k_{1}=0$, this leads to
the expression in Eq. \ref{nn3}. 

So far we have given only one relationship between $k_{1}$ and $k_{2}$.
We need one more relation connecting the two vectors to determine them. In
this paper we use the difference between the exact and Thomas-Fermi kinetic
energies for the  ground and the excited-state configurations as the second
relation.  We now explain this.  We take 
\begin{equation}
k_{1}=C\left(3\pi^{2}\rho({\bf r})\right)^{\frac{1}{3}}
\label{6}
\end{equation}
where $C$ is a constant.  Thus at each point in the inhomogeneous electron
gas, the inner radius of the shell in k-space is determined by the density
at that point with the outer radius being given via Eq. \ref{4} as
\begin{equation}
k_{2} = \left((1+C^{3})3\pi^{2}\rho({\bf r})\right)^{\frac{1}{3}}\:;
\label{n2}
\end{equation}
$C=0$ of course corresponds to the ground-state. Now with a given set
of occupied and virtual orbitals for a given system, we fix C for an
excited-state configuration by demanding that the corresponding 
Thomas-Fermi kinetic energy, given by Eq. \ref{n1} with $C>0$ for 
the excited-state, have the same error as it does for the ground-state 
(evaluated with $C=0$). In the example of $Be^{2+}$ given above, $C=1.4$ 
gives an error of about $9.5\%$ for the the $2s3p$ configuration. Thus it 
is this value of $C$ that we shall use to evaluate the LDA exchange-energy 
and the corresponding potential in the self-consistent Kohn-Sham calculation
for the $2s3p$ configuration. We note that this is one possible way of fixing 
the value of $C$; better ways of doing so may also exist.  However, as we
show below, the value of $C$ determined in this manner works quite well
for the majority of excited-states investigated.

We have performed self-consistent Kohn-Sham calculations for excited-states
where the electrons from the innermost orbitals are excited (as pointed out above, 
the simplest distribution of wavevectors that we have taken represents precisely 
such an excited-state) within the local-spin-density approximation (LSDA) by 
taking the LSDA functional as
\begin{equation}
E_{x}^{LSDA}[\rho^{\alpha},\rho^{\beta}] = \frac{1}{2}E_{x}^{LDA}
[2\rho^{\alpha}] + \frac{1}{2}E_{x}^{LDA}[2\rho^{\beta}]
\label{7}
\end{equation}
The resulting exchange energy functional and the corresponding potential
have structure similar to the ground-state LSDA functional but with
a different coefficient given in terms of $C$.  For LSDA calculations
we need two different $C$s, one for each spin. As discussed earlier, these
are fixed by keeping the error in the Thomas-Fermi kinetic energy the same 
for both the ground and the excited states. By performing these calculations
within the exchange-only, we show that for the excited states values closer
to the $\Delta$-SCF Hartree-Fock excitation energies are obtained with 
non-zero $C$.  Further, for a given system, $C$ increases as one goes to higher 
excited states. 

Shown in table I are the excitation energies of helium. We show the energies 
for three different excited states ($2s2p\;^{3}P$), ($2p^{2}\;^{3}P$) and 
($2s3p\;^{3}P$) of the helium atom calculated with $C=0$ (i.e. the ground-state 
LSDA) and with $C$ determined as described above.  Since all three excited-states
have only up spin electrons, $C$ shown in the table corresponds to up spin. These 
are all states that can be represented by a single Slater-determinant so that the 
LSDA is expected to work well \cite{ziegler,vonB} for them. We compare our 
results with the exact $\Delta$-SCF results of Hartree-Fock theory. In all the
excited states considered, it is seen that whereas the error in the excitation
energy obtained from the regular LSDA is about $3 eV$, with the proposed functional
it only a fraction of an $eV$.  Thus it is clear that non-zero value of $C$ gives a
better value for the excitation energy. Further, for the higher excited states the 
value of $C$ is larger, although it is slightly smaller when one goes from 
($2s2p\;^{3}P$) to ($2p^{2}\;^{3}P$).

To further check the validity of our approach, we have also tested it on
excited states of other systems. Shown in table II are the excitation energies 
of the $(2s2p\;^{3}P)$ state of the $Li^{+}$ ion, $(2s3p\;^{3}P)$ state of the 
$Be^{2+}$ ion, $(2p^{3}\;^{4}S)$ state of $Li$ atom and the 
$(1s^{1}2s^{2}2p^{1}\;^{3}P)$ state of the $Be$ atom. In the first three of 
these states, up spin electrons are promoted to higher energy orbitals so the $C$ 
given is that for the up spin.  For the $Be$ atom, the down spin electrons of $1s$ 
state is flipped and promoted to the $2p$ level.  Thus it is the down spin electron
in the $2s$ state that has to be described by a shell in the k-space; thus $C$ in 
this system is that for the down spin.  Further, since in this case both up and 
down spin electrons are involved, $C$ is fixed so that the error in the total 
Thomas-Fermi kinetic energy matches for the ground- and the excited-states.
We again see that for non-zero positive values of $C$, determined with the 
prescription given above, excitation energies come out to be closer to the 
$\Delta$-SCF Hartree-Fock excitation energies than with $C=0$.

Shown in table III are the numbers for the fluorine atom and neon positive
ion excited-states.  One of the excited-states ($1s^{1}2s^{2}2p^{6}\;^{2}S$) in 
each system corresponds to a shell in the k-space, whereas the other one 
($1s^{2}2s^{1}2p^{6}\;^{2}S$) does not.  As pointed out earlier, a shell in 
k-space represents well an excited-state in which the lowest lying orbitals are 
vacant.  Thus we see that for the ($1s^{1}2s^{2}2p^{6}\;^{2}S$) state of both the 
systems, the error in the excitation energy as given by the proposed functional is 
smaller by a factor of about two in comparison to the corresponding error in the 
LSDA excitation energy. The relative error in the case of LSDA is about $1.2\%$ 
whereas our functional gives an error of $0.6\%$. Since the excitation in these
cases involve single-electron being transferred, TDDFT calculation can also be 
performed to determine the excitation energy. For the fluorine atom, TDDFT gives
the excitation energy from the ground to the ($1s^{1}2s^{2}2p^{6}\;^{2}S$) to
be $23.7848$ a.u. which is in error by $29.94$ eV.  Similarly for the neon
ion, the excitation energy comes out to be $29.9615$ a.u. which is in error
by $33.48$ eV. 

The other excited-states($1s^{2}2s^{1}2p^{6}\;^{2}S$) shown in table III are 
those in which the lowest lying orbitals are not vacant, since one of the $2s$ 
electrons has been excited to the 
$2p$ orbital in fluorine or in a mono-positive ion of neon. In these cases, the 
corresponding wavevectors will not form a shell but will 
be distributed in some other manner; one possibility is an occupied sphere (of
radius $k_{1}$) representing the core states, then a vacant shell (from radius
$k_{1}$ to $k_{2}$) for the unoccupied states followed again by an occupied shell 
(from $k_{2}$ to $k_{3}$) representing the outer electrons. Thus the functional of 
Eq. \ref{5} above is not expected to be as accurate for such excited-states 
as  it is for those with empty lowest states; although it should still be better 
than the ground-state LSDA. This is clear from numbers in table III:  we see that
although the error in the excitation energy does become smaller, but not as much 
as in the cases discussed earlier. Also the relative error in these case is quite
large. Work on functionals with a different k-space occupation, which is more 
appropriate for such excited-states, is in progress. TDDFT calculations in these
cases give quite accurate energies \cite{harb1}. 

We have shown above that the correct local-density approximation for the
kinetic energy and  exchange energy in excited-state density-functional 
theory \cite{levy1} is made by taking the electrons to be occupying wavevectors 
differently than for the ground-state, and gives results that are superior 
to those obtained by applying the same approximation for the ground as well 
as the excited states.  As an example of this, we took the simplest case in 
which the occupied wave-vectors form a spherical shell. Through this we have
demonstrated that if consistency is maintained in making the LDA for different 
states, the resulting excitation energies are much better than those obtained by 
employing the same LDA for the ground- and the excited-state. Although our
method of finding the inner and out radii is an $ad-hoc$ one, some justification
for it exists on the basis of conjointness \cite{march} of the kinetic and
exchange energy.

To understand the functional proposed by us better, we have also looked at 
the spherical average \cite{gunnar} of the Fermi-hole when the k-space is 
occupied differently than for the ground-state.  We find that the spherical 
averaged hole corresponding to the functional proposed is closer to the 
spherically-averaged exact hole than that corresponding to the ground-state 
functional. These results will be presented in the future.

Next question that we address is if the excited-state LDA proposed by us 
can be generalised to include the gradient corrections.  The answer is
in the affirmative. As the first step, we assume that the LDA functional
is changed but the gradient corrections are the same for both the ground-
and excited-state functionals. Further, in this paper we have focussed on 
kinetic and exchange energy. The correlation energy could also be better 
approximated using a similar approach. Work in these directions has already
been started.  

The motivation for the present work stems from the requirement of excited-state 
density-functional theory that functionals for excited states be state-dependent.
The result of making the functionals state-dependent is that the errors in the 
total energy for both the ground- and excited-states are roughly the same and hence
the difference of energies comes out to be accurate in comparison to the results
of Hartree-Fock theory, exact-exchange calculations \cite{levy1} using the 
optimized potential method \cite{talman} or the near exact exchange
calculations \cite{deb} using the Harbola-Shani potential \cite{harb2}.
We note that in the latter three theories, the state-dependence of the exchnage 
functional is automatically accounted for by the use of exchnage-energy functional 
that depends on orbitals rather than the density.

{\bf Acknowledgement:} We thank Professor K.D. Sen for providing data on
different excited-states of atoms. We also thank Professor R. Prasad for
fruitful discussion.

\newpage
\begin{table}
\caption{Total energies and the excitation energies (in atomic units) of 
three different
excited-states of helium atom for $C=0$ (ground-state functional) and
the  value of $C$ determinded by comparison of exact and Thomas-Fermi
kinetic energies for up spin.  The corresponding Hartree-Fock (HF) 
excitation energies determined by $\Delta$-SCF method are given in the 
last line of each set. The last coulmn gives in eV the magnitude of the 
difference between the HF and the DFT transition energies.}
\vspace{0.2in}
\begin{tabular}{ccccc} 
\hline                                                            
State & C & Total Energy(a.u.) &   Excitation Energy(a.u.) &   Error (eV)\\ 
\hline
 ${\bf He\; atom}$ & & & \\
$1s^{2}(^{1}S)$ & - & -2.7236  & - & - \\
$2s2p(^{3}P)$ & 0.0  & -0.7223 & 2.0014 & 2.90 \\
           & 1.045 & -0.6095 & 2.1141 & 0.16 \\
           &      && $\Delta E_{HF}=$ 2.1081 &\\
\\
$2p^{2}(^{3}P)$ & 0.0  & -0.6965 & 2.0271 & 3.62 \\
           & 0.955 & -0.5933 & 2.1303 & 0.81 \\
           &      && $\Delta E_{HF}=$ 2.1603 &\\
\\
$2s3p(^{3}P)$ & 0.0  & -0.5615 & 2.1621 & 3.47\\
           & 1.395  & -0.4646 & 2.2590 & 0.83 \\
           &     & &$\Delta E_{HF}=$ 2.2898 &\\
\hline
\end{tabular}
\end{table}

\newpage

\begin{table}
\caption{Total energies and the excitation energies (in atomic units) of an
excited-state of lithium ion $Li^{+}$, berrylium ion $Be^{2+}$ and 
lithium atom for $C=0$ and the value of $C$ determinded by  comparison 
of the exact and Thomas-Fermi kinetic energies.  The corresponding 
Hartree-Fock excitation energy is given in the last line. The last coulmn 
gives in eV the magnitude of the difference between the HF and the DFT 
transition energies.}
\vspace{0.2in}
\begin{tabular}{ccccc} 
\hline                                                            
State & C & Total Energy(a.u.) &   Excitation Energy(a.u.) &   Error (eV)\\ 
\hline
 ${\bf Li^{+}\; ion}$ & & & & \\
$1s^{2}(^{1}S)$ & - & -7.0086  & - & - \\
$2s2p(^{3}P)$ & 0.0  & -1.8228 & 5.1858 & 4.89 \\
           & 1.06  & -1.6361 & 5.3725 & 0.19 \\
           &      && $\Delta E_{HF}=$ 5.3655 &\\
  ${\bf Be^{2+}\; ion}$ & & & & \\
$1s^{2}(^{1}S)$ & - & -13.2943  & - & - \\
$2s3p(^{3}P)$ & 0.0  & -2.5488 & 10.7455 & 8.28 \\
  & 1.421  & -2.3253 & 10.9691 & 2.19  \\
  &      &&$\Delta E_{HF}=$ 11.0499 &\\
 ${\bf Li\; atom}$ & & & &\\
$1s^{2}2s(^{2}S)$ & - & -7.1934  & - & - \\
$2p^{3}(^{4}S)$ & 0.0  & -2.1061 & 5.0873 & 7.32  \\
  & 0.777  & -1.9262 & 5.2672 & 2.43  \\
  &      &&$\Delta E_{HF}=$ 5.3565 &\\
  ${\bf Be\; atom}$ & & & & \\
$1s^{2}2s^{2}(^{1}S)$ & - & -14.2233  & - & - \\
$1s^{1}2s^{2}2p^{1}(^{3}P)$ & 0.0  & -10.1470 & 4.0863 & 3.07\\
  & 1.062  & -10.0582 & 4.1646 & 0.94 \\
  &      &&$\Delta E_{HF}=$ 4.1991 &\\
\hline
\end{tabular}
\end{table}

\begin{table}
\caption{The table caption is the same as that for Table II except that
the numbers are for fluorine atom and neon ion.}
\vspace{0.2in}
\begin{tabular}{ccccc} 
\hline                                                            
State & C & Total Energy(a.u.) &   Excitation Energy(a.u.) &   Error (eV)\\ 
\hline
  ${\bf F\; atom}$ & & & & \\
$1s^{2}2s^{2}2p^{5}(^{2}P)$ & - & -98.4740  & - & - \\
$1s^{1}2s^{2}2p^{6}(^{2}S)$ & 0.0  & -73.9002 & 24.5738 & 8.47\\
  & 0.685  & -73.4263 & 25.0477 & 4.42 \\
  &      &&$\Delta E_{HF}=$ 24.8852 & \\
\\
$1s^{2}2s^{1}2p^{6}(^{2}S)$ & 0.0  & -97.8069 & 0.6671 & 5.74\\
  & 0.238  & -97.7492 & 0.7248 & 4.17 \\
  &      &&$\Delta E_{HF}=$ 0.8781 & \\
  ${\bf Ne^{+}\; ion}$ & & & \\
$1s^{2}2s^{2}2p^{5}(^{2}P)$ & - & -126.7371  & - \\
$1s^{1}2s^{2}2p^{6}(^{2}S)$ & 0.0  & -95.8931 & 30.8440 & 9.47 \\
  & 0.670  & -95.3537 & 31.3834  & 5.20 \\
  &      &&$\Delta E_{HF}=$ 31.1921 & \\
\\
$1s^{2}2s^{1}2p^{6}(^{2}S)$ & 0.0  & -125.9027 & 0.8344 & 6.76 \\
  & 0.244  & -125.8311 & 0.9060  & 4.81 \\
  &      &&$\Delta E_{HF}=$ 1.0829 & \\
\hline
\end{tabular}
\end{table}

\end{document}